\documentclass[10pt,aps,prb,twocolumn,singlespace]{revtex4}
\pdfoutput=1
\usepackage{amsmath}
\usepackage{graphicx}
\usepackage{subfig}
\usepackage{setspace}

\everymath{\displaystyle}

\begin{document}
\title{Optimized Effective Potential Using The Hylleraas Variational Method}
\author{T.W. Hollins}
\email{t.w.hollins@durham.ac.uk}
\affiliation{Department of Physics, Science Laboratories, University of Durham, Science Labs, South Road, Durham, DH1 3LE, UK.}
\author{S.J. Clark}
\email{s.j.clark@durham.ac.uk}
\homepage{cmt.dur.ac.uk/sjc}
\affiliation{Department of Physics, Science Laboratories, University of Durham, Science Labs, South Road, Durham, DH1 3LE, UK.}
\author{K. Refson}
\email{keith.refson@stfc.ac.uk}
\affiliation{STFC, Rutherford Appleton Laboratory, Didcot, Oxfordshire, OX11 0QX, UK.}
\author{N. Gidopoulos}
\email{nikitas.gidopoulos@stfc.ac.uk}
\affiliation{STFC, Rutherford Appleton Laboratory, Didcot, Oxfordshire, OX11 0QX, UK.}

\begin{abstract}
In electronic structure calculations the optimized effective potential
(OEP) is a method that treats exchange interactions exactly using a
local potential within density-functional theory (DFT). We present a
method using density functional perturbation theory combined with the
Hylleraas variational method for finding the OEP by direct minimization which avoids any sum over unoccupied
states. The method has been implemented within the plane-wave,
pseudopotential formalism. Band structures for zinc blende
semiconductors Si, Ge C, GaAs, CdTe and ZnSe, wurtzite semiconductors
InN, GaN and ZnO and the rocksalt insulators CaO and NaCl have been
calculated using the OEP and compared to calculations using the local
density approximation (LDA), a selection of generalized gradient
approximations (GGAs) and Hartree-Fock (HF) functionals. The band gaps
found with the OEP improve on the calculated results for the LDA, GGAs
or HF, with calculated values of 1.16eV for Si, 3.32eV for GaN and
3.48eV for ZnO. The OEP energies of semi-core d-states are also
greatly improved compared to LDA.

\end{abstract}

\maketitle

\section{Introduction}

Density functional theory (DFT) is one of the great successes in
tackling the many-electron
problem\cite{hohnbergkohn64,kohnsham,KohanoffGidopoulos}.
Density-dependent exchange-correlation potentials, such as the local
density approximation (LDA) or generalized gradient approximations
(GGAs), are useful and surprisingly accurate for many properties of
materials despite the known deficiencies of these
functionals. Generally, the LDA slightly underestimates lattice
constants, while GGAs slightly overestimate them and both
underestimate single-particle band gaps in
solids.\cite{JonesGunnarssonreview,FilippiSinghUmrigar,ShamSchluter,PerdewLevy}
The exchange energy can be calculated
exactly using the non-local Hartree-Fock (HF) formalism, rather than
being approximated as part of the LDA or GGA. However pure HF grossly
overestimates band gaps in solids and is considerably more expensive
computationally because of the fully non-local nature of the exchange
operator.

The central theorem of density functional theory\cite{hohnbergkohn64} demonstrates that
a local potential $V(\mathbf{r})$ is sufficient to completely define a many-electron system in the ground state.
A local potential that includes the effect of the exact
exchange interaction was first proposed by Sharp and
Horton\cite{sharphorton} and solved nearly a quarter of century later
by Talman and Shadwick.\cite{talmanshadwick} This potential is known
as the optimized effective potential (OEP), because a local potential
is created to optimally represent
the non-local HF exchange potential. The total energy of the
OEP system is variational with the potential in the same way the LDA or GGA total energy is
variational in the electronic density. This allows the construction of
equations to describe the potential within a Kohn-Sham DFT formalism.
The resultant single particle excitation energies agree much better with experiment
than those calculated using the LDA, GGAs and HF.\cite{KohanoffGidopoulos}

The improved OEP description of excited states is a consequence of the
freedom from self-interaction error
of both occupied and unoccupied Kohn-Sham states\cite{KummelKronikreview}. 
Within the LDA and GGAs 
the potential experienced by all states includes some degree of 
self-interaction\cite{perdewzunger}. In HF the exchange term cancels 
the self-interaction error in the occupied states, but no such cancellation
is present for the unoccupied states. 
This results in too small an electronic band gap using the
LDA and GGAs and too
large in HF\cite{GarboKreibichKurthGross,kummelperdew, salagorling}.

There is a distinction between
the fundamental band gap of a material and the corresponding
Kohn-Sham gap.  The Kohn-Sham gap ($E^{KS}_g$) is
defined as the difference between the eigenvalues of the highest
occupied orbital and the lowest unoccupied orbital
\begin{equation}
E^{KS}_{g} = \epsilon_{N+1} - \epsilon_{N},
\end{equation}
where \begin{math} \epsilon_{i} \end{math} is the eigenvalue of the
$i$th orbital and N is the total number of electrons in the
system. Whereas the fundamental band gap ($E_g$) is defined as the
difference in the ionization potential and the electron affinity
\begin{equation}
E_{g} = I-A = E(N+1) - 2E(N) + E(N-1),
\end{equation}
where $I$ is the ionization potential, $A$ is the electron affinity
and $E(N)$ is the total energy of a $N$ electron system. The two are
related by
\begin{equation}
E_{g} = E^{KS}_{g} + \Delta_{xc},
\end{equation}
where $\Delta_{xc}$ is the derivative discontinuity in XC energy with
respect to particle
number.\cite{PerdewLevy,ShamSchluter,ShamSchluter2} This constant is
the energy associated with a change in the exchange-correlation
potential when a infinitesimal charge is added to a $N$ electron
system.  While a comparison between the Kohn-Sham and fundamental band
gaps would be instructive, the value of the derivative
discontinuity is unknown for most real materials. (See however Godby 
\emph{et al}\cite{GodbySchluterSham} and Gorling \emph{et al.}\cite{StadeleGorling1,StadeleGorling2})


In finite systems, the exchange potential should decay as
-1/r in the long-range limit for all states. In the OEP this decay
occurs for all states irrespective of occupancy. In HF the non-local potential
for occupied states behaves correctly but for the empty states decays
exponentially and in the LDA/GGA the potential for all states decays
exponentially\cite{GGA-exp}.  Furthermore, the orbitals in the OEP each correctly
decay with an individual exponent, whereas in HF all occupied orbitals
decay with the same exponent.
\cite{KohanoffGidopoulos,GarboKreibichKurthGross,KummelKronikreview,kummelperdew,
  salagorling}

The OEP is potentially as versatile as LDA/GGA and HF methods.
It could be used instead of HF as a foundation
for orbital-dependent potentials (so called hybrid
functionals)\cite{KohanoffGidopoulos}. Furthermore once the local
potential has been obtained, the calculation of system properties is
faster than HF based methods, as the computationally expensive application 
of the exchange operator is avoided.

While the OEP correctly describes many system properties (see Grabo
\emph{et al}\cite{GarboKreibichKurthGross}), 
it is not as widely used as the LDA, GGA and HF-hybrid
based methods. This can be attributed in part to the lack of comparably efficient 
and robust computational schemes to evaluate it.  In most
formulations of the OEP method a sum over all excited states
is required and truncation to a finite sum yields a slowly-convergent series. 
Furthermore an accurate description of these high energy states requires large 
basis sets at high computational expense if a local basis description is used.
Despite these deficiencies, some calculations have been done in
solids within the linear muffin-tin orbital (LMTO) basis
set\cite{kotani1,kotani2,kotani3}, the FPLAPW method\cite{Gorling3} as well as in plane-wave
pseudopotential
implementations\cite{BylanderKleinman1,BylanderKleinman2,BylanderKleinman3,StadeleGorling1,StadeleGorling2}.
The agreement of the calculated results in the previous work above
with experiment can be very good (see Ref.[\onlinecite{KummelKronikreview}]). However most applications have been
to semiconductors. For wide-gap insulating systems the performance of the OEP
can be poor, notably for noble-gas solids.\cite{MagyarFleszarGross}
See Ref.~[\onlinecite{KummelKronikreview}] for full details and further discussion of the OEP method
and its results.

The extremely demanding task of calculating the full OEP can be
reduced by using the mean-field approximation of Krieger, Li and
Iafrate\cite{KLI} (well-known as the KLI approximation) and even in
this less precise formalism impressively
accurate results have been obtained\cite{BylanderKleinman1,BylanderKleinman2,BylanderKleinman3,FukaAkai}.

The principal aim of this work is to derive and demonstrate a
variational method of calculating the full exchange-only OEP
without the
need for a sum over all unoccupied states of a system. 
To this
end, our method for the calculation of the OEP is formulated using
density functional perturbation theory (DFPT) and the Hylleraas variational
principle.  As in the case phonon or electric field
perturbations, only occupied Kohn-Sham orbitals are explicitly
included.  This method is then applied to a range of semiconductors
and insulators.

\section{Theory}
\subsection{The Optimized Effective Potential}



We first summarize the OEP before deriving a variational
implementation within the density functional perturbation formalism.
For the remainder of this article we consider only the exchange-only OEP and will simply use the term OEP from here on.
A starting point for finding the local exchange only OEP is the
non-local Hartree-Fock equation
\begin{align} 
& \left[ \hat{T}(\mathbf{r}) + \hat{V}_{\mathrm{ext}}(\mathbf{r}) + \hat{V}_\mathrm{H}(\mathbf{r}) + \hat{V}^\sigma_\mathrm{X}(\mathbf{r}) \right] \phi^\sigma_i(\mathbf{r})  = \epsilon^\sigma_i \phi^\sigma_i(\mathbf{r}),
\end{align} 
where \begin{math} \hat{T}(\mathbf{r}) \end{math} is the kinetic energy, \begin{math} \hat{V}_\mathrm{ext}(\mathbf{r}) \end{math} is the external potential, \begin{math} \hat{V}_\mathrm{H}(\mathbf{r}) \end{math} is the Hartree potential, \begin{math} \hat{V}_\mathrm{X}(\mathbf{r}) \end{math} is the HF exchange potential, \begin{math} i \end{math} indexes the electronic states, \begin{math} \sigma \end{math} is the spin index and \begin{math} \phi^\sigma_i(\mathbf{r})\end{math} is the $i$th orbital with spin $\sigma$.
The Kohn-Sham (KS) equation\cite{hohnbergkohn64,kohnsham} that incorporates the OEP is given by
\begin{equation} 
\left[ \hat{T}(\mathbf{r}) + \hat{V}_{\mathrm{ext}}(\mathbf{r})+ \hat{V}^\sigma(\mathbf{r}) \right] \phi^\sigma_i(\mathbf{r}) = \epsilon^\sigma_i \phi^\sigma_i(\mathbf{r}),
\label{eq. KohnSham}
\end{equation} 
where \begin{math}{\hat{V}^\sigma}\end{math} is the effective
potential fulfilling the role of both the Hartree and exchange
potentials. The usual derivation of the OEP uses a chain rule
expansion of the derivative of the exchange-correlation energy with
respect to the density\cite{GarboKreibichKurthGross}, however here a
treatment inspired by perturbation theory is used.

For a potential which differs from the ground state (GS) potential by an amount
$\delta V^\sigma(\mathbf{r})$ such that
\begin{math}{V^\sigma(\mathbf{r})=V^\sigma(\mathbf{r})+\delta
    V^\sigma(\mathbf{r})}\end{math} the GS KS orbitals change by
\begin{equation} 
\phi^\sigma_i(\mathbf{r}) \rightarrow \phi^\sigma_i(\mathbf{r}) + \int d\mathbf{x} \delta V^\sigma(\mathbf{x}) \frac{\delta \phi^\sigma_i(\mathbf{r})}{\delta V^\sigma(\mathbf{x})},
\end{equation} 
where from first order perturbation theory\cite{gonze} the last term
of the above equation becomes
\begin{align} 
\frac{\delta \phi^\sigma_i(\mathbf{r})}{\delta V^\sigma(\mathbf{x})} = & -\sum^{N^\sigma}_{j=1}\frac{\phi^\sigma_j(\mathbf{r}) \phi^{\sigma*}_j(\mathbf{x})}{\epsilon^\sigma_j - \epsilon^\sigma_i} \phi^\sigma_i(\mathbf{x}) \nonumber \\     
& -\sum^{\infty}_{a=N^\sigma+1}\frac{\phi^\sigma_a(\mathbf{r}) \phi^{\sigma*}_a(\mathbf{x})}{\epsilon^\sigma_a - \epsilon^\sigma_i} \phi^\sigma_i(\mathbf{x}).
\label{eq: 1st order}
\end{align} 
The change in the effective potential also gives a first order change
in the total energy, thus
\begin{equation} 
\triangle E[V^\sigma] = \int d\mathbf{x} \delta V^\sigma(\mathbf{x}) \frac{\delta E[V^\sigma]}{\delta V^\sigma(\mathbf{x})}.
\end{equation} 


The OEP energy
$E[V^\sigma]$ is given by the Hartree-Fock functional evaluated using the Kohn-Sham OEP orbitals. Using
the chain rule and the perturbation of the orbitals, the derivative of
the energy with respect to the potential is given by
\begin{align} 
\frac{\delta E[V^\sigma]}{\delta V^\sigma(\mathbf{x})} = & \int d \mathbf{r} \sum^{N^\sigma}_{i=1}\frac{\delta E[V^\sigma]}{\delta \phi^\sigma_i(\mathbf{r})} \frac{\delta \phi^\sigma_i(\mathbf{r})}{\delta V^\sigma(\mathbf{x})} + \mathrm{h.c.}.
\label{eq: chainrule}
\end{align}
By applying the Hellmann-Feynman theorem (for brevity the explicit $\mathbf{r}$ dependence of the potentials has been omitted) 
\begin{align} 
\frac{\delta E[V^\sigma]}{\delta \phi^\sigma_i(\mathbf{r})} = \left[ \hat{T} + \hat{V}_{\mathrm{ext}}+\hat{V}_\mathrm{H}+\hat{V}^\sigma_\mathrm{X} \right] \phi^{\sigma*}_i(\mathbf{r}),
\label{eq: hellman}
\end{align}
and substituting equations (\ref{eq: 1st order}), (\ref{eq: hellman}) into equation (\ref{eq: chainrule}), we obtain
\begin{align} 
\frac{\delta E[V^\sigma]}{\delta V^\sigma(\mathbf{x})} = & -\int d \mathbf{r} \sum^{N^\sigma}_{i=1} \sum^{\infty}_{a=N^\sigma+1} \frac{\phi^\sigma_a(\mathbf{r}) \phi^{\sigma*}_a(\mathbf{x})\phi^\sigma_i(\mathbf{x})}{\epsilon^\sigma_a - \epsilon^\sigma_i} \nonumber \\
& \times \left[ \hat{T} + \hat{V}_{\mathrm{ext}}+\hat{V}_\mathrm{H}+\hat{V}^\sigma_\mathrm{X} \right] \phi^{\sigma*}_i(\mathbf{r})+ \mathrm{h.c.}.
\end{align}
By substitution of the Kohn-Sham equation (eq. \ref{eq. KohnSham}) the above can be written as\cite{yangwu}
\begin{align} 
\frac{\delta E[V^\sigma]}{\delta V^\sigma(\mathbf{x})} = & - \sum^{N^\sigma}_{i=1}\sum^{\infty}_{a=N^\sigma+1} \left[ \frac{ \langle \phi^{\sigma}_i | \hat{V}_\mathrm{H} + \hat{V}^\sigma_\mathrm{X} - \hat{V}^\sigma | \phi^{\sigma}_a \rangle}{\epsilon^\sigma_a - \epsilon^\sigma_i} \right. \nonumber \\
\Big. & \times \phi^{\sigma*}_a(\mathbf{x}) \phi^{\sigma}_i(\mathbf{x}) \Bigg] + \mathrm{h.c.},
\label{OEPequation}
\end{align}
which is known as the OEP equation and was first derived by Sharp and
Horton\cite{sharphorton}.  
Here it is important to note that the
requirement to sum over all unoccupied states presents a challenge for
convergence of the system properties. A very large number of unoccupied
orbitals must be included to ensure adequate convergence and
consequently the calculation is extremely
demanding\cite{GorlingHeblmannJonesLevy}. The truncation of the
infinite sum has also been the subject of
concerns about the analytic properties of the
solutions\cite{GidopoulosLathiotakis}.

\subsection{Applying The Hylleraas Variational Principle}

We present the following formalism to evaluate the OEP which avoids an explicit
sum over states by borrowing ideas from density functional perturbation 
theory\cite{Baroni,gonze} using the Hylleraas variational method. 
The effective potential is
obtained by variational minimization of $E[V^\sigma (\mathbf{r})]$. The
minimization direction for the potential is defined by the functional
derivative of the energy with respect to the effective potential
\begin{equation} 
V^\sigma (\mathbf{r}) \rightarrow V^\sigma (\mathbf{r}) - \lambda \frac{\delta E[V^\sigma]}{\delta V^\sigma(\mathbf{r})}.
\end{equation} 

The functional derivative of the energy can alternatively be written
as
\begin{equation} 
\frac{\delta E[V^\sigma]}{\delta V^\sigma(\mathbf{x})} = \sum^{N^\sigma}_{i=1} \phi^{\sigma}_i(\mathbf{x}) \left( \tilde{\phi}^{\sigma}_i(\mathbf{x}) \right)^* + \mathrm{h.c.},
\label{eq: E wrt V}
\end{equation}
where
\begin{equation} 
\tilde{\phi}^{\sigma}_i(\mathbf{x}) = -\sum^{\infty}_{a=N^\sigma+1} \frac{ \phi^{\sigma}_a(\mathbf{x}) \langle \phi^{\sigma}_a | \hat{V}_\mathrm{H} + \hat{V}^\sigma_\mathrm{X} - \hat{V}^\sigma | \phi^{\sigma}_i \rangle}{\epsilon^\sigma_a - \epsilon^\sigma_i},
\label{eq: 1st order wave}
\end{equation}
and \begin{math} \tilde{\phi}^{\sigma}_i(\mathbf{x}) \end{math} is the
first order correction to an unperturbed orbital \begin{math}
  \phi^{\sigma}_i(\mathbf{x}) \end{math}. These first order corrections
to the orbitals were named ``orbitals shifts'' by K\"{u}mmel and
Perdew.\cite{kummelperdew,kummelperdew2}

An alternative method of calculating \begin{math}
  \tilde{\phi}^{\sigma}_i(\mathbf{x}) \end{math} without explicitly
including any unoccupied states is to use the the Hylleraas
variational principle\cite{Hylleraas}. We define a second order variational functional
\begin{align}
G^{\sigma}_i [\tilde{\phi}^{\sigma}_i] & = \langle \tilde{\phi}^{\sigma}_i |(\hat{T} + \hat{V}_{ext} + \hat{V}^\sigma - {\epsilon}^{\sigma}_i)| \tilde{\phi}^{\sigma}_i \rangle \nonumber \\
& + \langle \tilde{\phi}^{\sigma}_i |(\hat{V}_\mathrm{H} + \hat{V}^\sigma_\mathrm{X} - \hat{V}^\sigma - \tilde{\epsilon}^{\sigma}_i)| \phi^\sigma_{i} \rangle \nonumber \\
& + \langle \phi^\sigma_{i} |(\hat{V}_\mathrm{H} + \hat{V}^\sigma_\mathrm{X} - \hat{V}^\sigma - \tilde{\epsilon}^{\sigma}_i)| \tilde{\phi}^{\sigma}_i \rangle,
\label{eq: Hylleraas_G}
\end{align}
where \begin{math} \tilde{\epsilon}^{\sigma}_i \end{math} is the first order correction to the eigenvalues. 
$G^{\sigma}_i [\tilde{\phi}^{\sigma}_i]$ is also variational with respect to $\tilde{\phi}^{\sigma}_i$ 
under the constraint
\begin{equation} 
\langle \phi^{\sigma}_i | \tilde{\phi}^{\sigma}_i \rangle + \langle \tilde{\phi}^{\sigma}_i |\phi^{\sigma}_i \rangle =0
\end{equation}
and using the Hylleraas variational principle it can be shown that the orbital shifts which minimize the second-order functional also satisfy equation~(\ref{eq: 1st order wave}).
By substitution of the projection operator
\begin{equation} 
P_c = \hat{I} - \sum^{N^\sigma}_{j=1} | \phi^{\sigma}_j \rangle \langle \phi^{\sigma}_j | = \sum^{\infty}_{a=N^\sigma+1} | \phi^{\sigma}_a \rangle \langle \phi^{\sigma}_a |
\end{equation}
equation~(\ref{eq: Hylleraas_G}) may be rewritten
\begin{align}
G^{\sigma}_i [\tilde{\phi}^{\sigma}_i] & = \langle \tilde{\phi}^{\sigma}_i |(\hat{T} + \hat{V}_{ext} + \hat{V}^\sigma - {\epsilon}^{\sigma}_i)| \tilde{\phi}^{\sigma}_i \rangle \\
& + \langle \tilde{\phi}^{\sigma}_i |( \hat{I} - \sum^{N^\sigma}_{j\not = i} | \phi^{\sigma}_j \rangle \langle \phi^{\sigma}_j |) (\hat{V}_\mathrm{H} + \hat{V}^\sigma_\mathrm{X} - \hat{V}^\sigma)| \phi^\sigma_{i} \rangle \nonumber \\
& + \langle \phi^\sigma_{i} |(\hat{V}_\mathrm{H} + \hat{V}^\sigma_\mathrm{X} - \hat{V}^\sigma) ( \hat{I} - \sum^{N^\sigma}_{j\not=i} | \phi^{\sigma}_j \rangle \langle \phi^{\sigma}_j |)| \tilde{\phi}^{\sigma}_i \rangle. \nonumber 
\end{align}


The exact first order correction to the orbitals \begin{math} \tilde{\phi}^{\sigma}_i \end{math} that minimizes
\begin{math} G^\sigma_i[\tilde{\phi}^{\sigma}_i] \end{math}
can also be found using
the Sternheimer-like equation\cite{Sternheimer}
\begin{align} 
& (\hat{T} + \hat{V}_{ext} + \hat{V}^\sigma - \epsilon^{\sigma}_i ) | \tilde{\phi}^{\sigma}_i \rangle \nonumber \\ & + (\hat{I} - \sum^{N^\sigma}_{j=1} | \phi^{\sigma}_j \rangle \langle \phi^{\sigma}_j |) (\hat{V}_\mathrm{H} + \hat{V}^\sigma_\mathrm{X} - \hat{V}^\sigma) | \phi^{\sigma}_i \rangle =0.
\label{eq: sternheimer}
\end{align}
Equation~(\ref{eq: sternheimer}) can be solved using iterative methods\cite{Baroni}.
Replacing the infinite sum over states in 
equation~(\ref{eq: 1st order wave}) with an iterative procedure and considering only 
the occupied  Kohn-Sham subspace avoids the severe convergence difficulties of the former method.
Instead of determining of how many additional states to include in equation~(\ref{eq: 1st order wave}), at a computational 
cost increasing with their number, the convergence problem is transformed to one of 
determining  how many iterative cycles are required. Consequently a reliable 
solution may be  achieved at considerably lower expense than summing 
equation~(\ref{eq: 1st order wave}) directly.

By combining equations (\ref{eq: E wrt V}) and (\ref{eq: sternheimer}),
the OEP can be found variationally while the first order correction to
the orbitals is obtained by solving the Sternheimer equation
directly.



Equations (\ref{eq: E wrt V}) and
(\ref{eq: sternheimer}) are similar to those given by K\"{u}mmel and
Perdew\cite{kummelperdew,kummelperdew2}. Our method differs from theirs by
exploiting the variational character of $ E[V^\sigma]$  using the explicit gradient of 
equation~(\ref{eq: E wrt V}) to perform a direct minimization. Their method involves a 
self-consistency cycle with an update procedure for $v_{\textsc{XC}}$ using a KLI-like expression 
and involving a division by the density. Our direct minimization is conceptually 
simpler and, because of the variational principle is likely to be more
numerically robust.


\section{Implementation and Convergence}

\begin{figure}
\centering
\includegraphics[scale=0.80, trim = 60mm 25mm 50mm 15mm, clip]{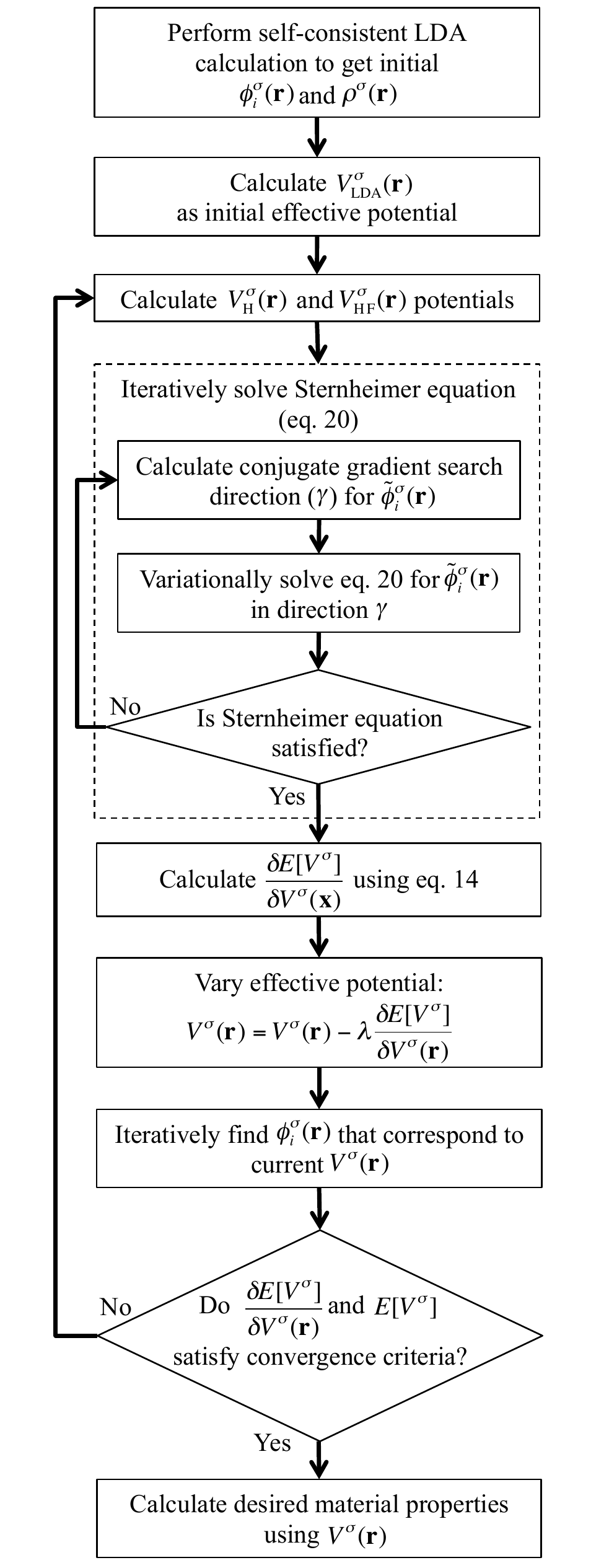}
\caption{\small Flow diagram showing procedure for finding the OEP.}
\end{figure}


Our procedure for finding the OEP is to solve a
double nested loop of minimizations. A flow diagram illustrating the
full procedure is shown in Fig. 1. The inner of the two loops represents
the solution of 
the Sternheimer equation (equation (\ref{eq: sternheimer}))
to find the first order correction to the orbitals. The method
is derived from the Baroni
Green's function solver technique, employing a conjugate gradient
minimization scheme to find \begin{math}
  \tilde{\phi}^{\sigma}_i(\mathbf{r}) \end{math} in a fixed
potential \begin{math}
  V^{\sigma}(\mathbf{r}) \end{math}.\cite{Baroni,PickardMauri,YatesPickardMauri}


Using the first order correction to the orbitals and the orbitals
themselves, the derivative of the total energy with respect to the
potential can be found from Equation~(\ref{eq: E wrt V}). This gradient
is used in the direction-set methods to variationally optimize the potential, and
its magnitude (residual norm) provides an 
indication of the level of convergence of the effective potential. The
variation of the potential is accomplished via either one of two
methods; the first is a conjugate gradient scheme\cite{Payne}, the
second method is a modified steepest descent method, known as the
Barzilai-Borwein (BB) method, which requires knowledge of the gradient
at the current point and previous point only\cite{BB}.

After each step of varying the effective potential, Kohn-Sham orbitals
in this new, fixed potential are found non self-consistently. When the
gradient and difference in total OEP Kohn-Sham energy per step is
smaller than a predetermined threshold. the calculation is considered
to be converged.

The procedure is initialized by first solving the KS equation using a
local density-dependent exchange-correlation potential, in this case
the LDA. From this self-consistent calculation the initial orbitals
and density are obtained. Using this density the effective potential
is then constructed explicitly using the LDA functional for the
exchange and correlation. Solving the KS equation to find the ground
state orbitals is accomplished variationally using a conjugate
gradient method.\cite{Payne} The non-local HF exchange energy and the
Hartree energy are also initially calculated using the LDA orbitals.

The OEP has been implemented within the pseudopotential plane-wave
code, CASTEP\cite{CASTEP1,CASTEP2}.
The orbitals, density and potentials
are represented on rectilinear grids in the usual manner of a plane-wave DFT
implementation.\cite{CarParrinello}  Kohn-Sham orbitals are described on reciprocal space grid points $\mathbf{G}$
within a sphere bounded by
the cut-off wavevector,
 \begin{math} G_\mathrm{max} \end{math} and the density and potentials 
are nonzero within a sphere of radius \begin{math}
  2G_\mathrm{max} \end{math}. Therefore the grids in both real and
reciprocal space used to represent the density and potentials have
twice the dimensions of the grid used for the orbitals. The Hylleraas minimization scheme is performed explicitly on these real
space grids by direct variation, so that the effective basis used to represent
$V^{\sigma}(\mathbf{r})$ is the set of grid points $\lbrace \mathbf{G}\rbrace: |\mathbf{G}| \le 2 G_{\mathrm{max}}$.

Optimized norm-conserving pseudopotentials used in this work were generated
using the Opium code\cite{Opium} developed by Rappe \emph{et
al.}. The Hartree-Fock approximation was used and
the non-analytic behaviour of HF pseudopotentials was treated using the localization and optimization scheme of Al-Saidi, Walter and
Rappe\cite{alshadirappe}. 
Including exchange exactly in the pseudopotentials is required to
ensure accurate core-valence interaction within the current
formalism. Incorrect core-valence interaction has been shown to have
noticeable effects on calculated electronic
structures.\cite{engel1,engel2}
We found that calculations using pseudopotentials which include
d-states in the core typically predict band gaps 20-70\% larger
than if the d-states are treated as
valence.
Therefore semi core d-states were treated as valence
for As, Cd, Ga, Ge, In, Se, Te and Zn.
\begin{figure}
\centering
\includegraphics[scale=0.5, trim = 0mm 0mm 0mm 0mm, clip]{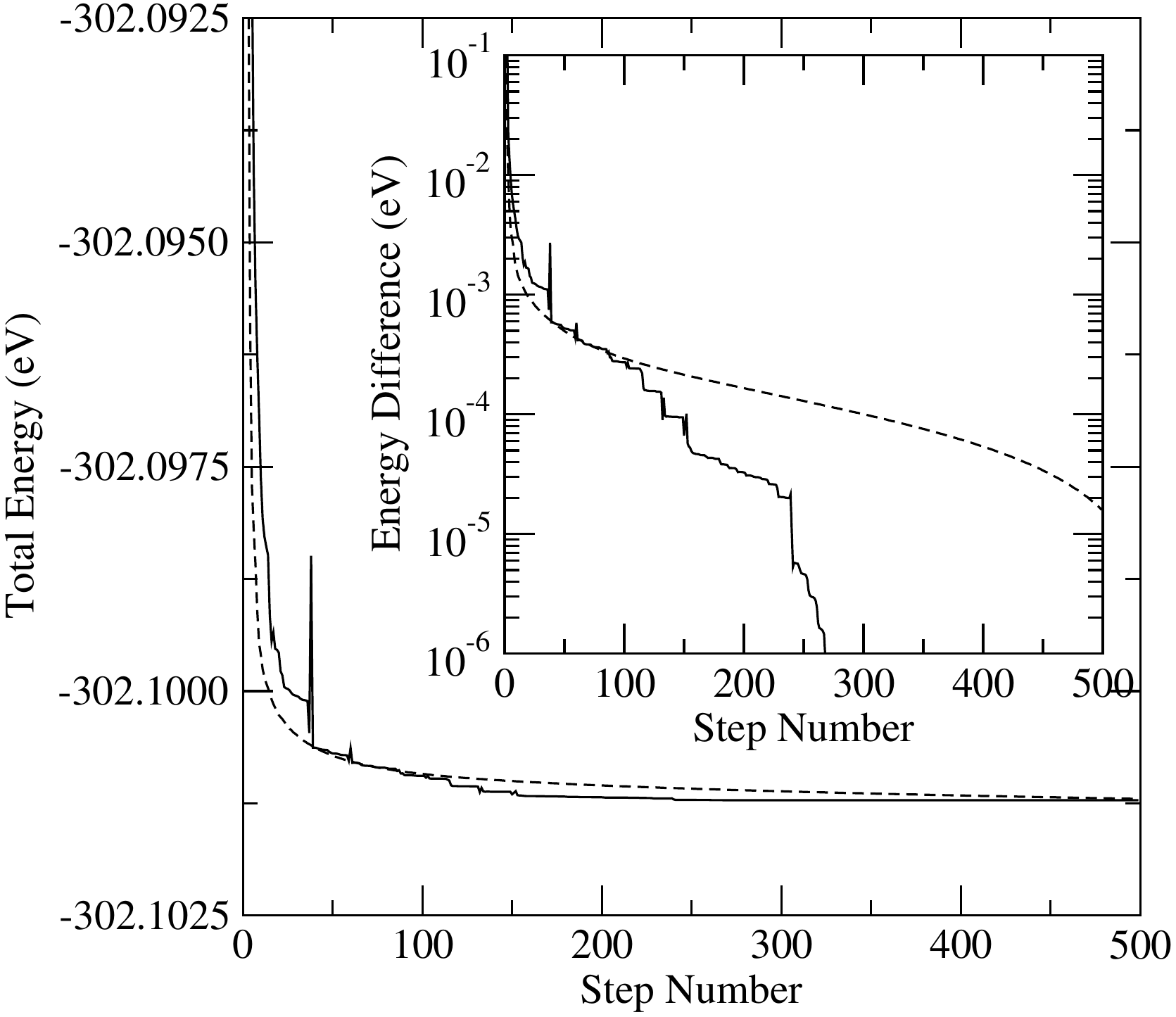}
\caption{\small Convergence of total Kohn-Sham energy for the OEP against number of steps. The solid line is the Barzilai-Borwein minimizer and dashed line for the conjugate gradient minimizer. Inset: Energy residual on a log-linear scale.}
\end{figure}

The convergence characteristics of the conjugate-gradient and Barzilai-Borwein variational minimizers, 
are compared in Figure 2 which  plots the total OEP energy for diamond as a function of iteration number.
An initial rapid decrease in the total energy is
followed by a long tail of decreasing change in the total energy in both cases.
The conjugate gradient energy decreases
smoothly and monotonically towards the converged result, but the Barzilai-Borwein
energy does not, exhibiting sharp drops and an occasional spiked increase (behaviour 
which has been noted previously \cite{BBreview}). 
After 100 or so steps the BB method still converges rapidly while the 
conjugate gradient method exhibits a very slow energy decrease.
The difference in total energy per step is
below 2.5$\mu$eV per atom after 250 iterations of the BB minimizer and
the solution is stationary after 300 iterations. For the conjugate
gradients method however the energy difference per step does not drop
below 5$\mu$eV per atom even after 500 iterations. After 500
iterations the energy difference between the two methods is within
8$\mu$eV per atom. In both cases we find that the calculated gradient
tends to zero as the total energy converges, as expected in a variational method.
The convergence rates of the $\Gamma$ point band gap of diamond behave in a similar fashion
as shown in Figure~3. After 250 iterations the
gap is converged to within 2.5$\mu$eV per atom for the BB
minimizer. By comparison, with the conjugate gradient minimizer the band
gap is converged to 50$\mu$eV per atom in the same number of iterations. 
All
calculations presented below using the OEP method were run for at
least 250 iterations to ensure sufficient accuracy.  


\begin{figure}
\centering
\includegraphics[scale=0.5, trim = 0mm 0mm 0mm 0mm, clip]{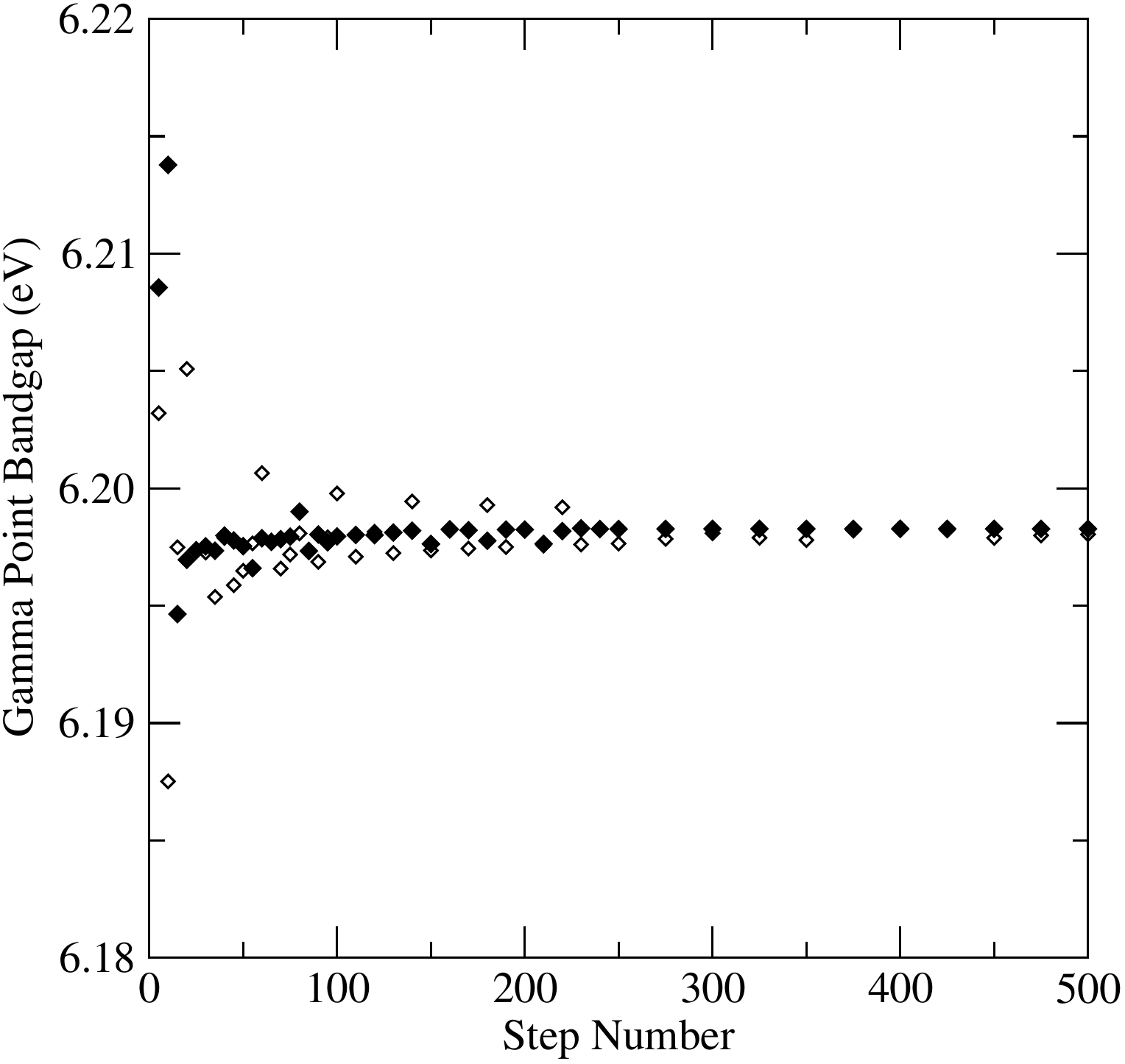}
\caption{\small Convergence of the $\Gamma$ point band gap of diamond
  using the OEP against number of steps. Solid dots are for the
  Barzilai-Borwein minimizer and open dots are for the conjugate
  gradient minimizer.}
\end{figure}

\begin{figure*}
\centering
\includegraphics[scale=0.50, trim = 12.5mm 22.5mm 12.5mm 45mm, clip]{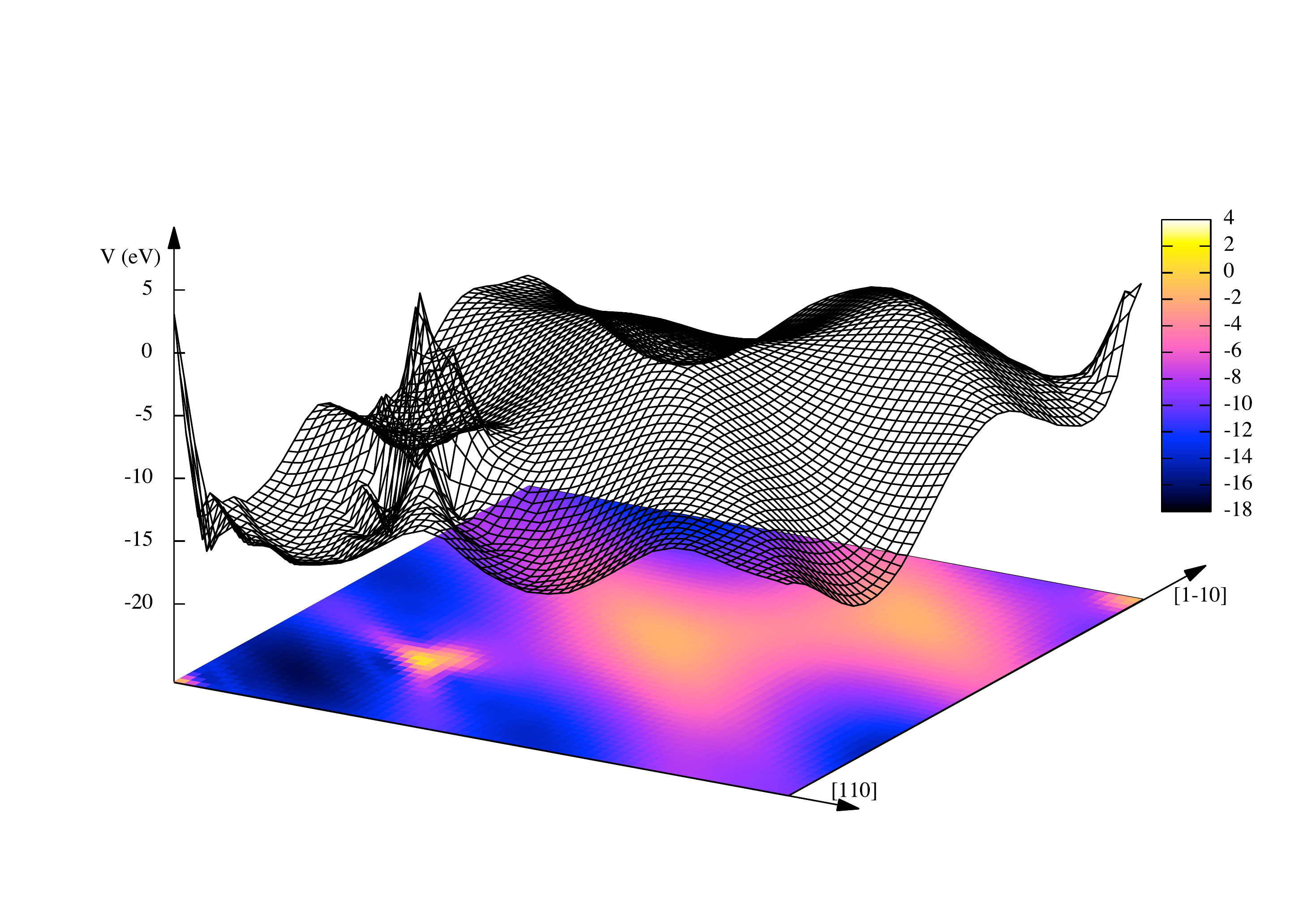}
\caption{\small The exchange potential ($V_\mathrm{OEP} -
  V_\mathrm{H}$) as a slice perpendicular to the [111] direction (bond
  axis) in diamond structured Si for the primitive unit cell. Atomic
  positions are at the origin and quarter of the long diagonal. The
  colour key is in eV.}
\label{fig: OEPpotential}
\end{figure*}

In the calculations that follow, the basis set size (plane-wave
cut-off energy) and Brillouin-Zone sampling were chosen so that total
energy differences, evaluated using the LDA, were less than 1
meV/atom.  The same settings were used for the OEP calculations, which
resulted in a very similar convergence error with cutoff as the LDA
case. However the OEP calculations were more sensitive to
Brillouin-Zone sampling error; for example, in the case of diamond,
$8\times 8\times 8$ resulted in a 1 meV/atom error using the LDA but 3
meV for OEP rising to 9 meV for ZnO. Convergence testing was also
performed on the FFT grid used to represent the potential and the
energy error was determined to be $\approx 10 \mu$eV.

\section{Calculated Electronic Properties}

An example OEP is displayed in figure \ref{fig: OEPpotential}, in this
case a slice through the primitive unit cell of Si in the
diamond structure. The potential is smoothly varying outside the
pseudopotential regions, which are the only regions where the
potential is positive. The potential is at its deepest in the bonding
region between the two ions and (outside the pseudopotential core
region) is always negative but tending to zero in the low density
regions away from the ions.

\begin{figure}
\centering
\includegraphics[scale=0.45, trim = 0mm 0mm 0mm 0mm, clip]{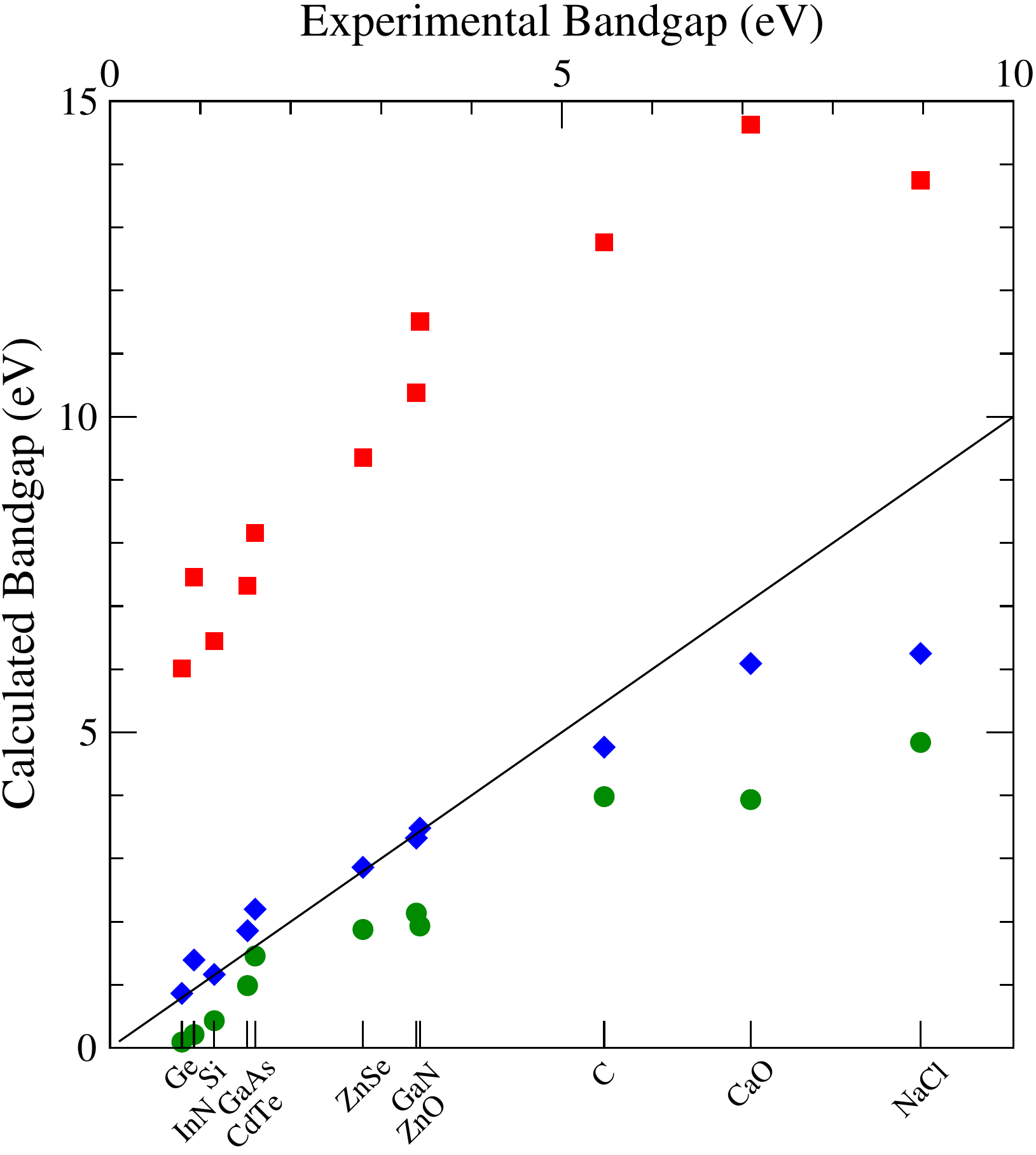}
\caption{\small Comparison of calculated and experimental band gaps,
  red squares are HF, blue diamonds are OEP and green circles are
  LDA. (Colour online)}
\label{fig: Linegraph}
\end{figure}

\begin{figure}
\centering
\subfloat{\includegraphics[scale=0.30, trim = 0mm 0mm 0mm 15mm, clip]{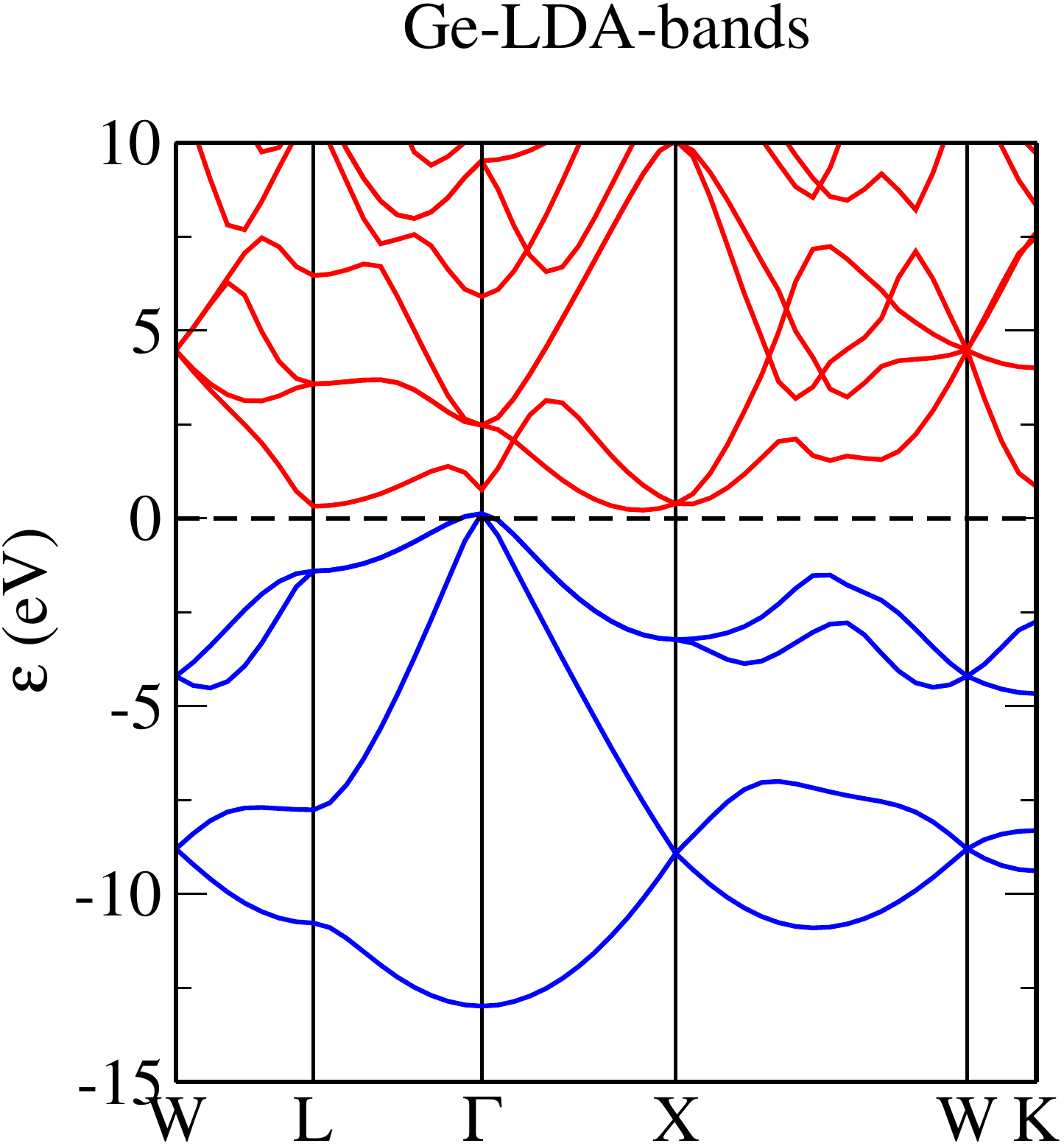}}
\subfloat{\includegraphics[scale=0.30, trim = 0mm 0mm 0mm 15mm, clip]{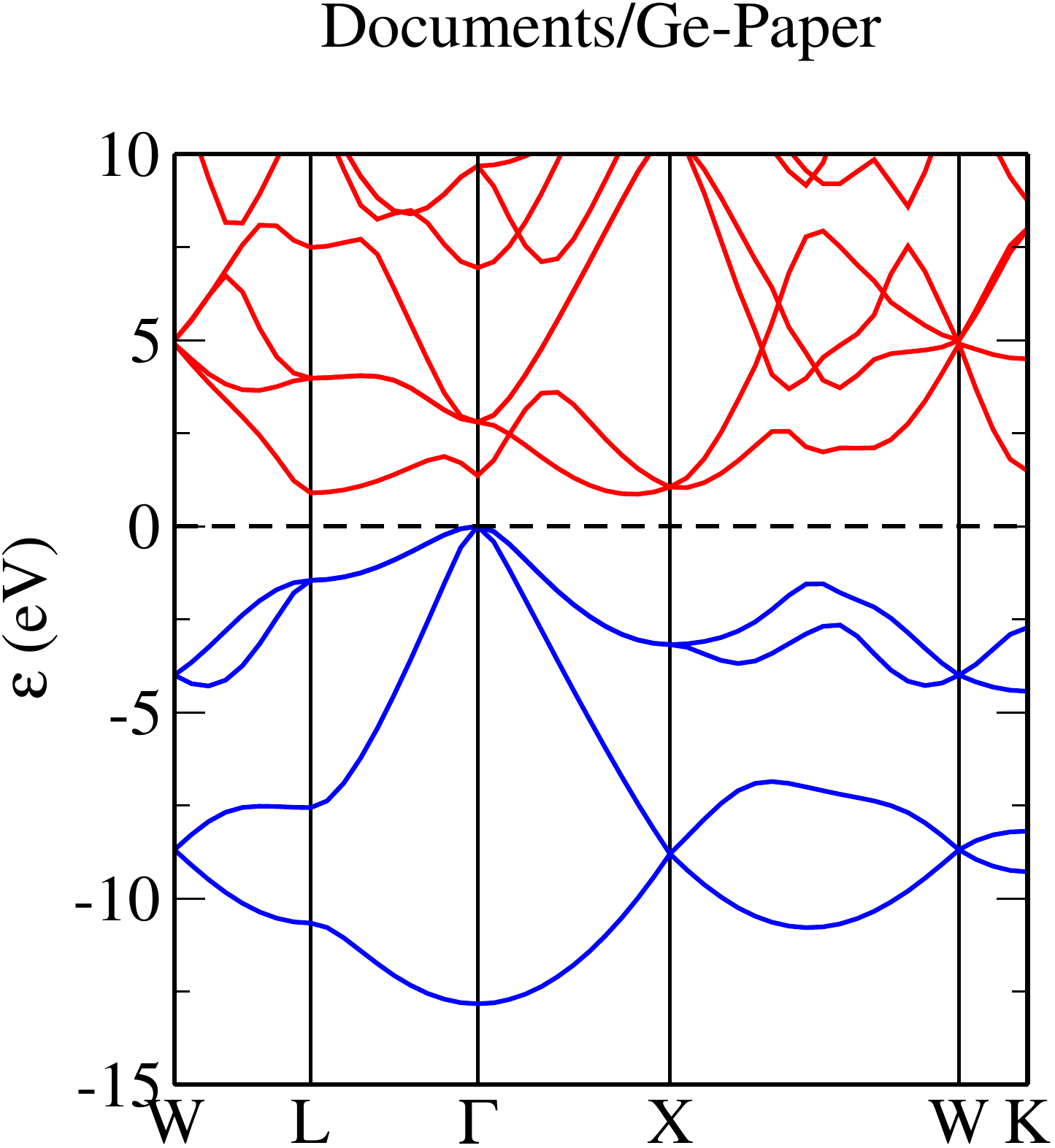}}
\caption{\small Band structure of Ge (diamond structure), calculated
  using the LDA (left) and the OEP (right).}
\label{fig: Ge}
\end{figure}

\begin{figure}
\centering
\subfloat{\includegraphics[scale=0.30, trim = 0mm 0mm 0mm 15mm, clip]{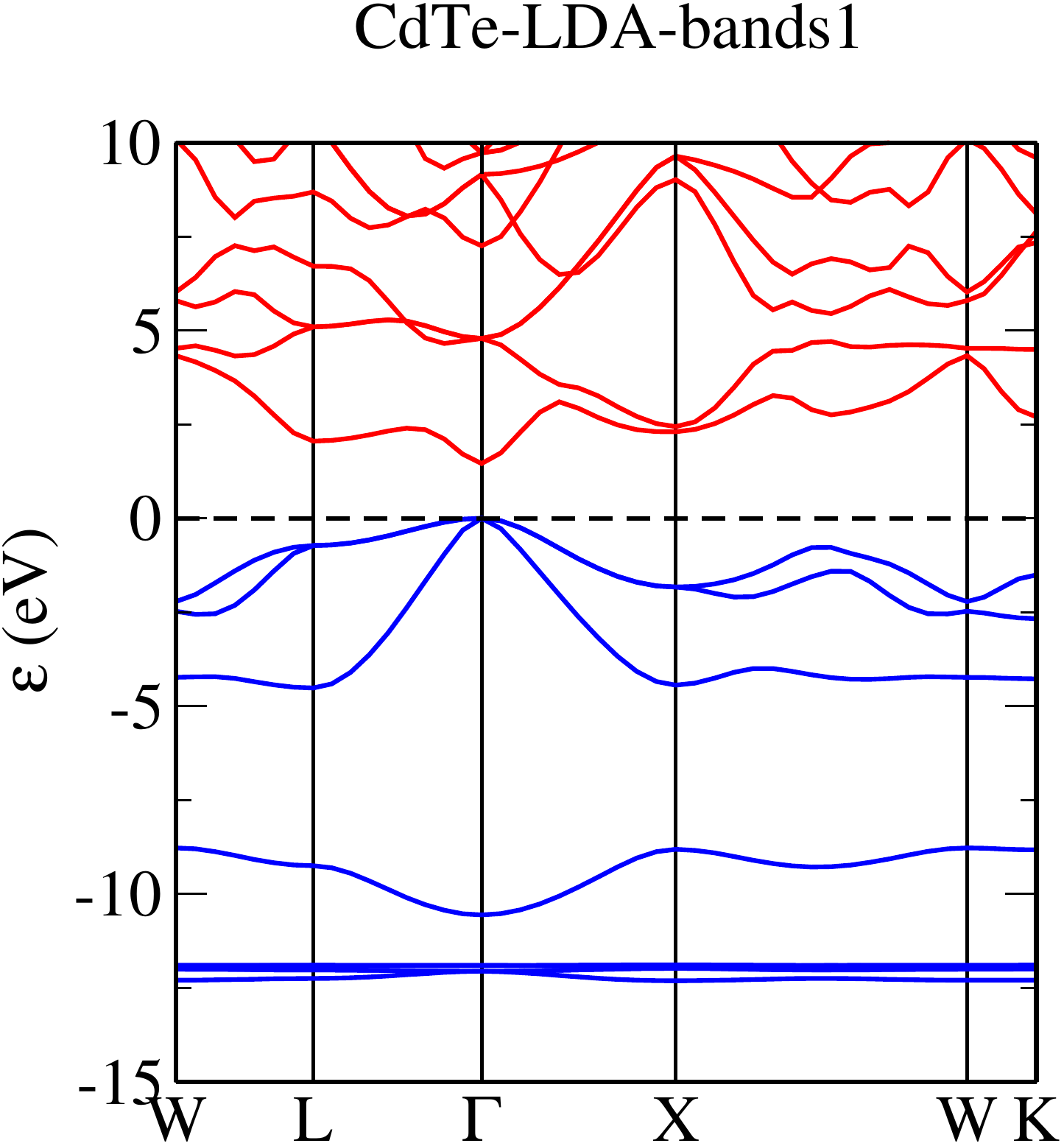}}
\subfloat{\includegraphics[scale=0.30, trim = 0mm 0mm 0mm 15mm, clip]{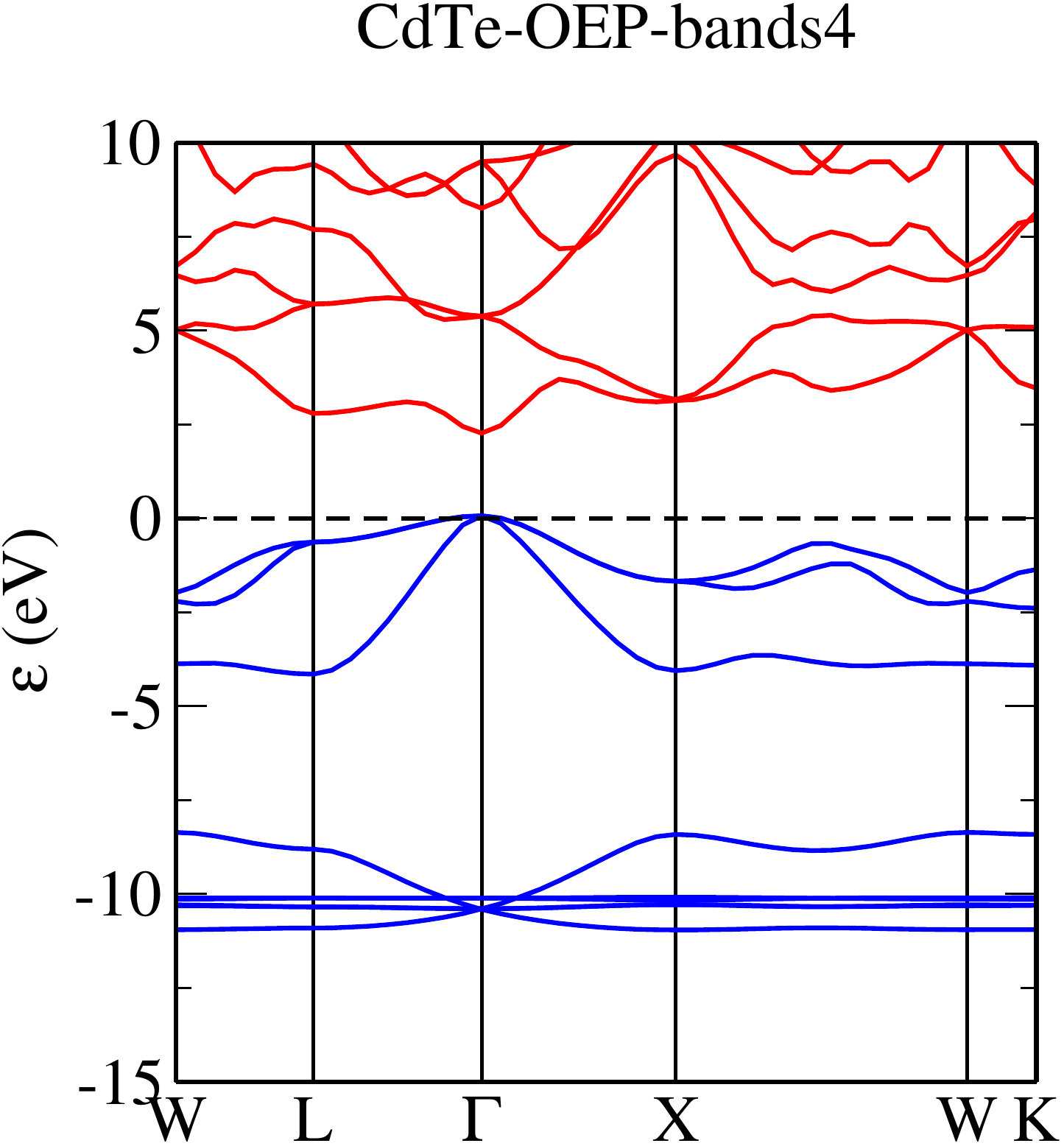}}
\caption{\small Band structure of CdTe (zinc blende structure),
  calculated using the LDA (left) and the OEP (right).}
\label{fig: CdTe}
\end{figure}

\begin{figure}
\centering
\subfloat{\includegraphics[scale=0.30, trim = 0mm 0mm 0mm 15mm, clip]{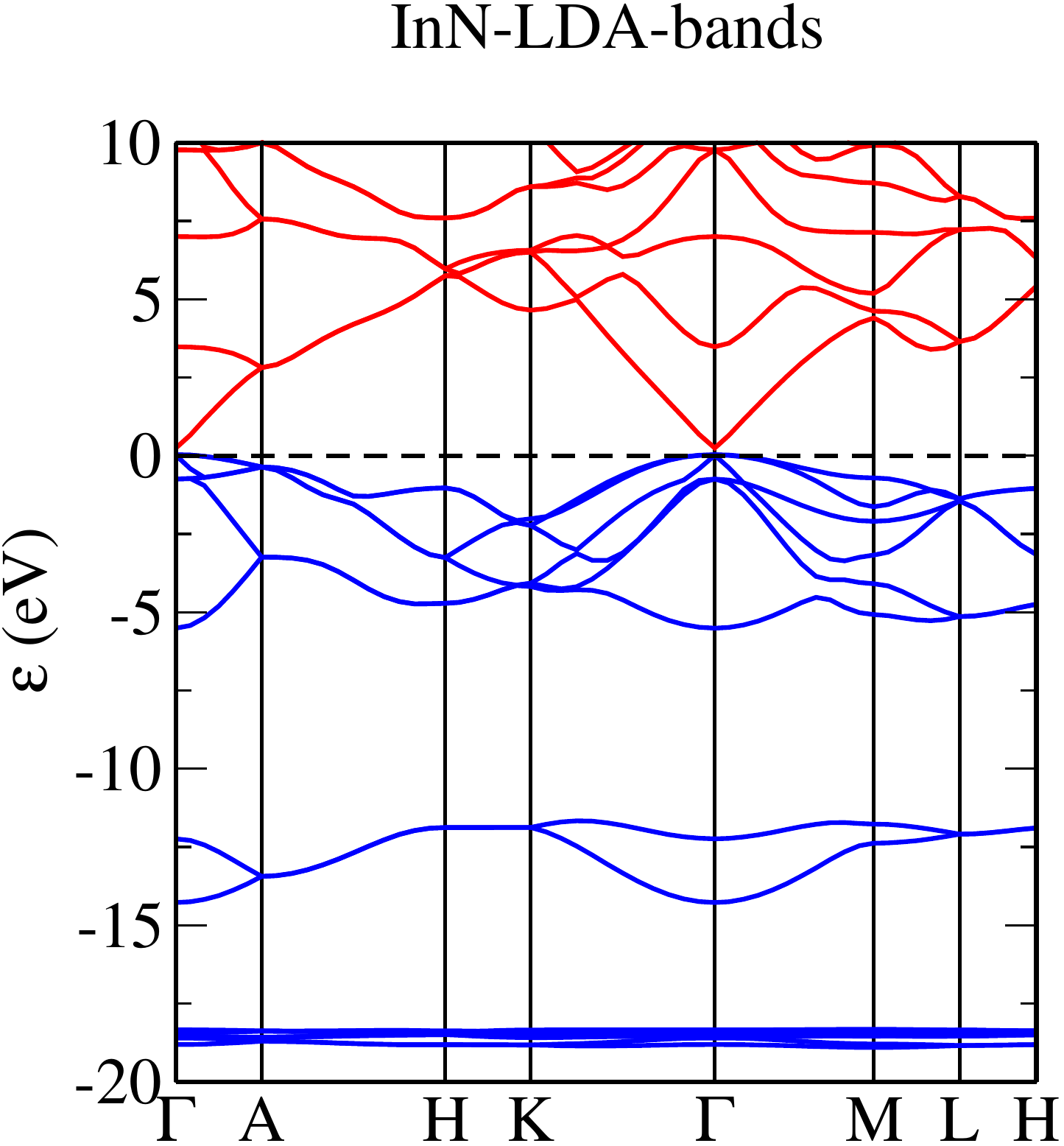}}
\subfloat{\includegraphics[scale=0.30, trim = 0mm 0mm 0mm 15mm, clip]{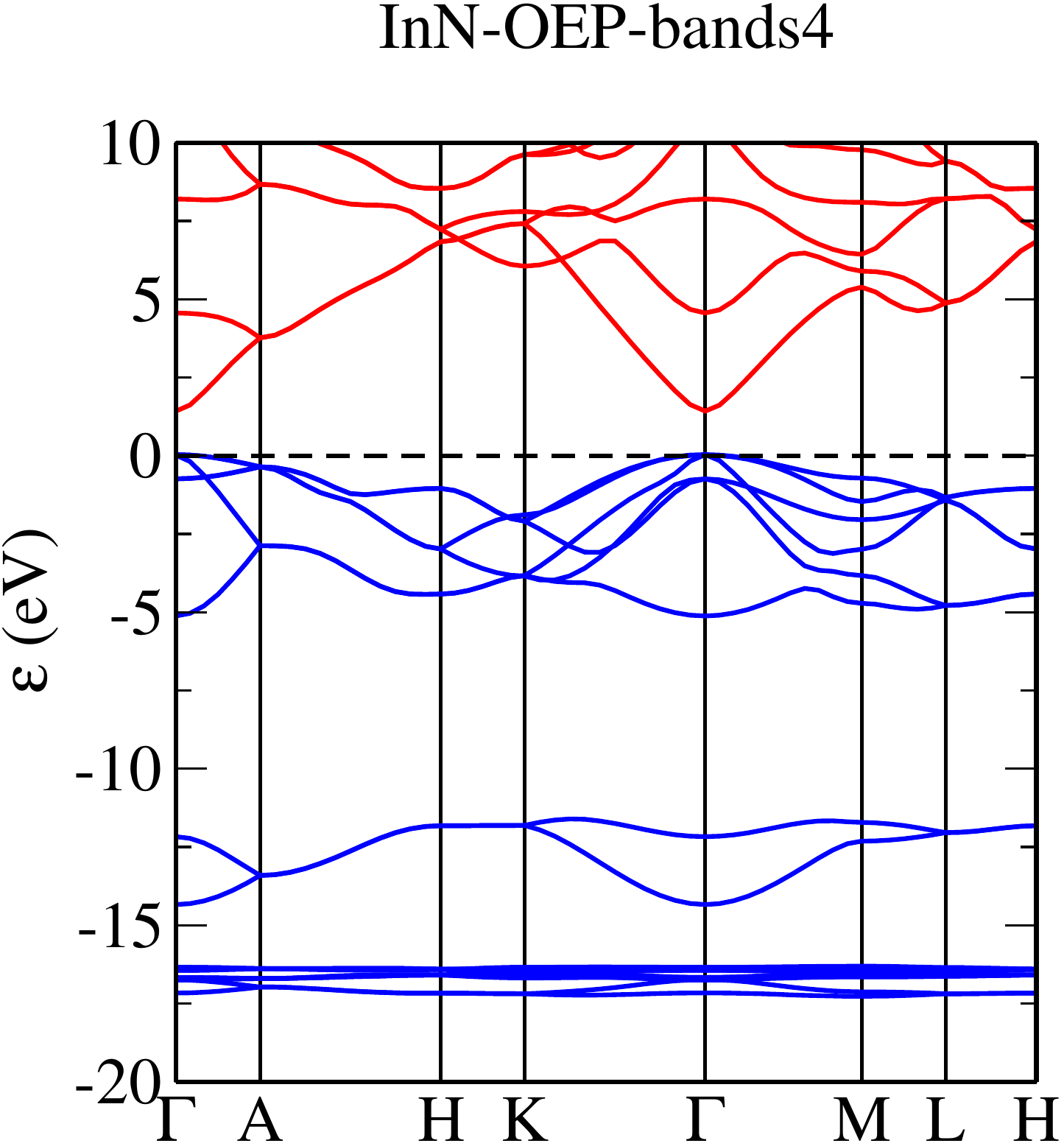}}
\caption{\small Band structure of InN (wurtzite structure) using the
  LDA (left) and the OEP (right).}
\label{fig: InN}
\end{figure}

\begin{figure}
\centering
\subfloat{\includegraphics[scale=0.30, trim = 0mm 0mm 0mm 15mm, clip]{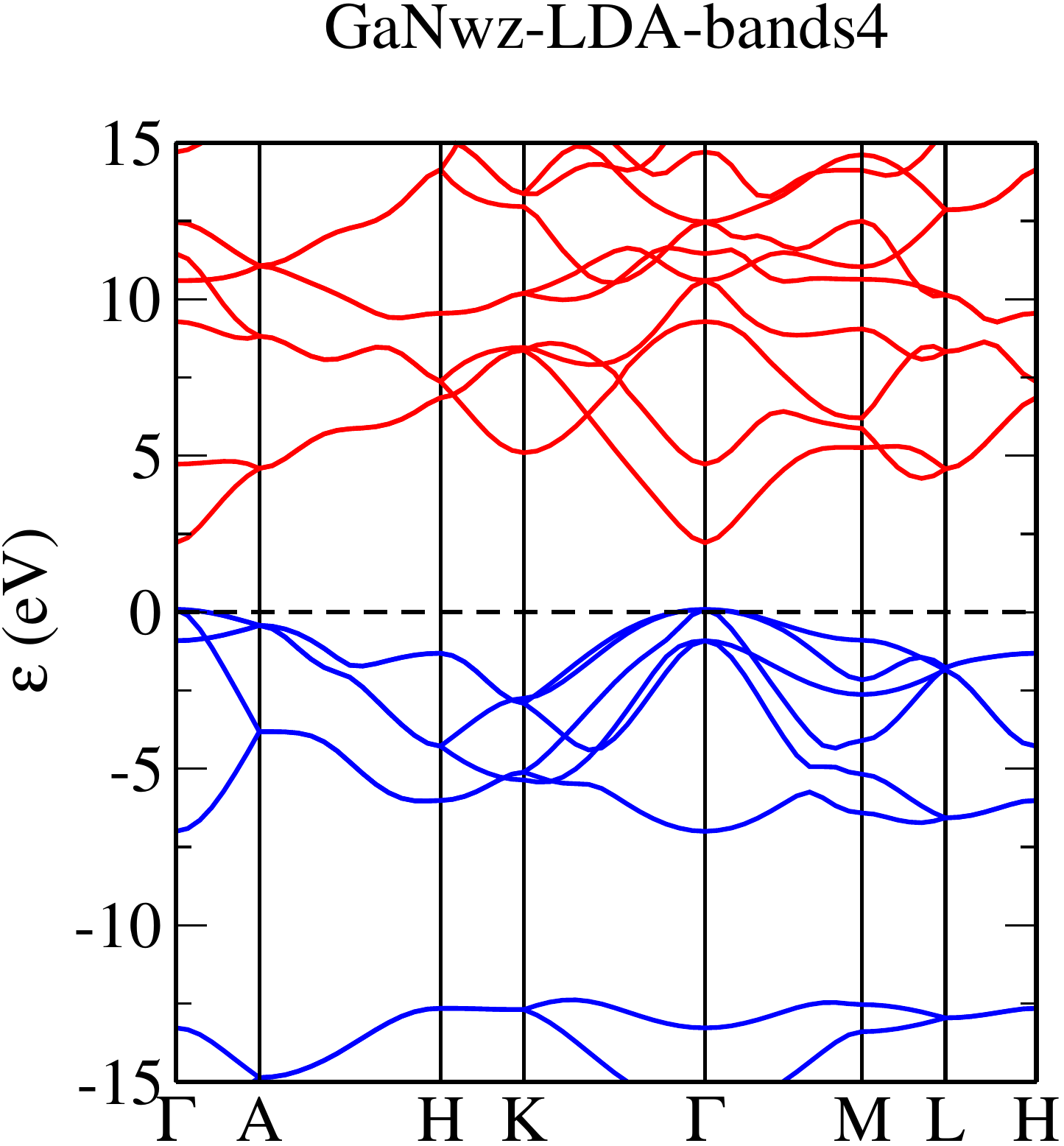}}
\subfloat{\includegraphics[scale=0.30, trim = 0mm 0mm 0mm 15mm, clip]{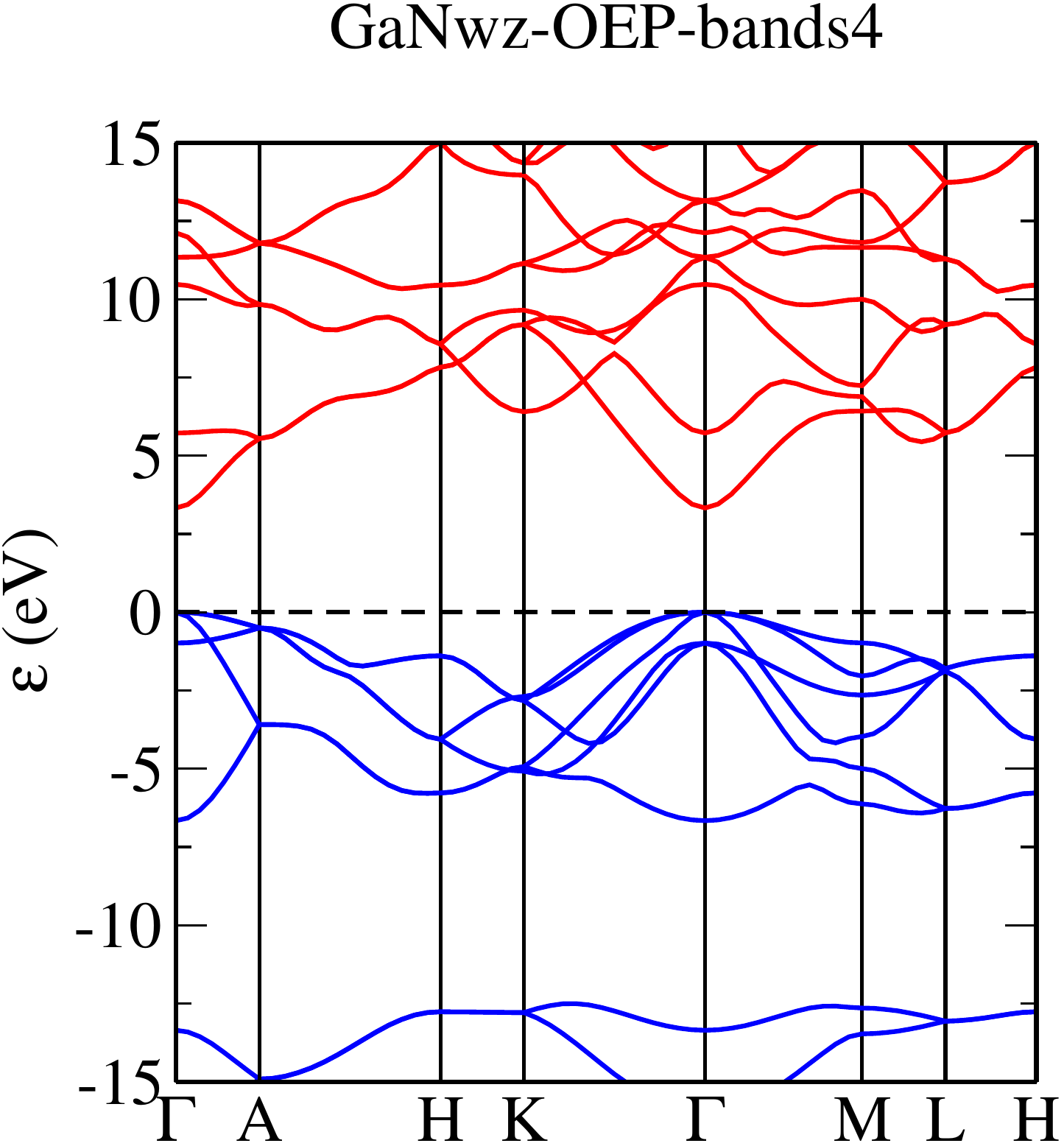}}
\caption{\small Band structure of GaN (wurtzite structure) using the
  LDA (left) and the OEP (right).}
\label{fig: GaN}
\end{figure}

\begin{figure}
\centering
\subfloat{\includegraphics[scale=0.30, trim = 0mm 0mm 0mm 15mm, clip]{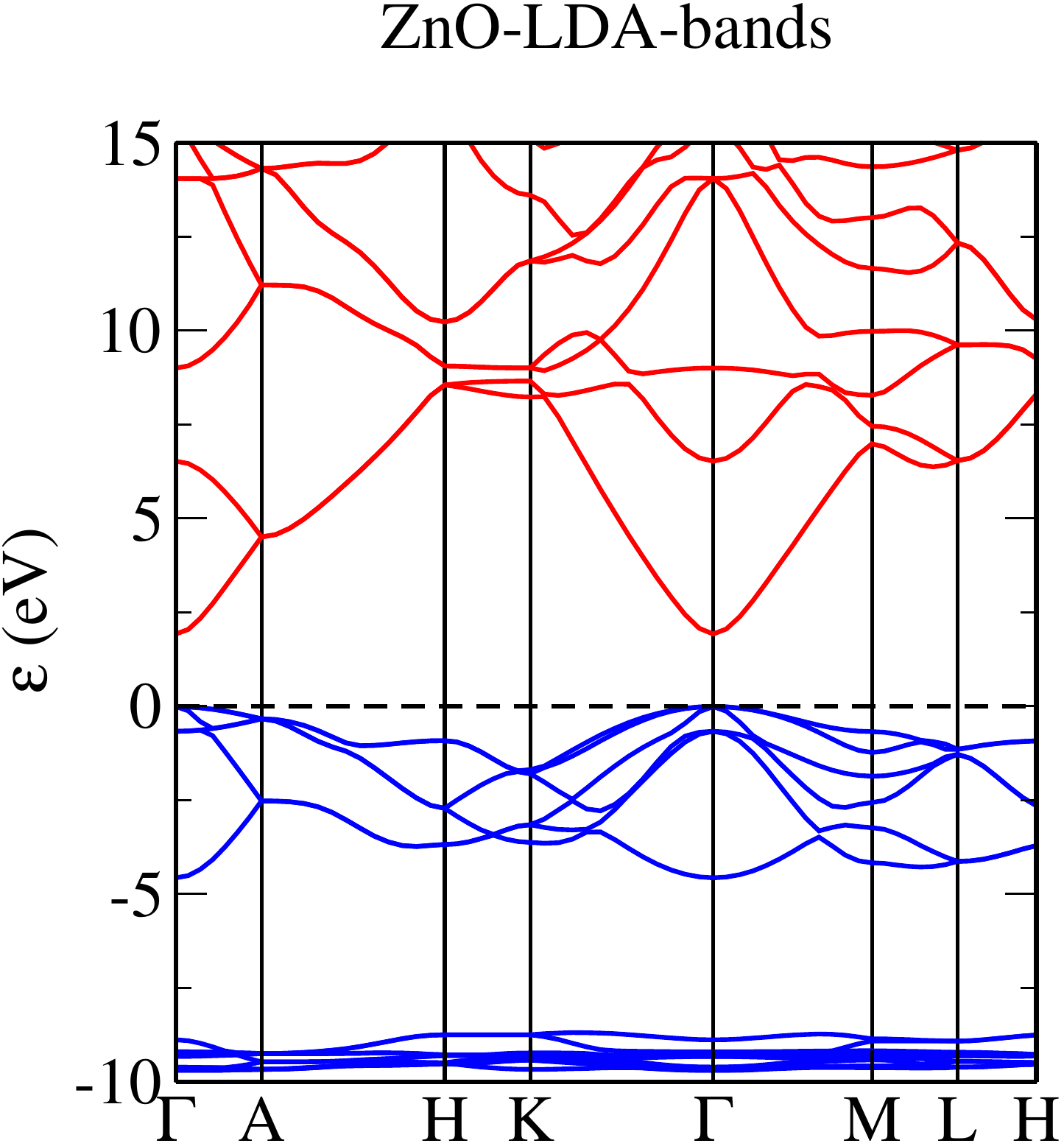}}
\subfloat{\includegraphics[scale=0.30, trim = 0mm 0mm 0mm 15mm, clip]{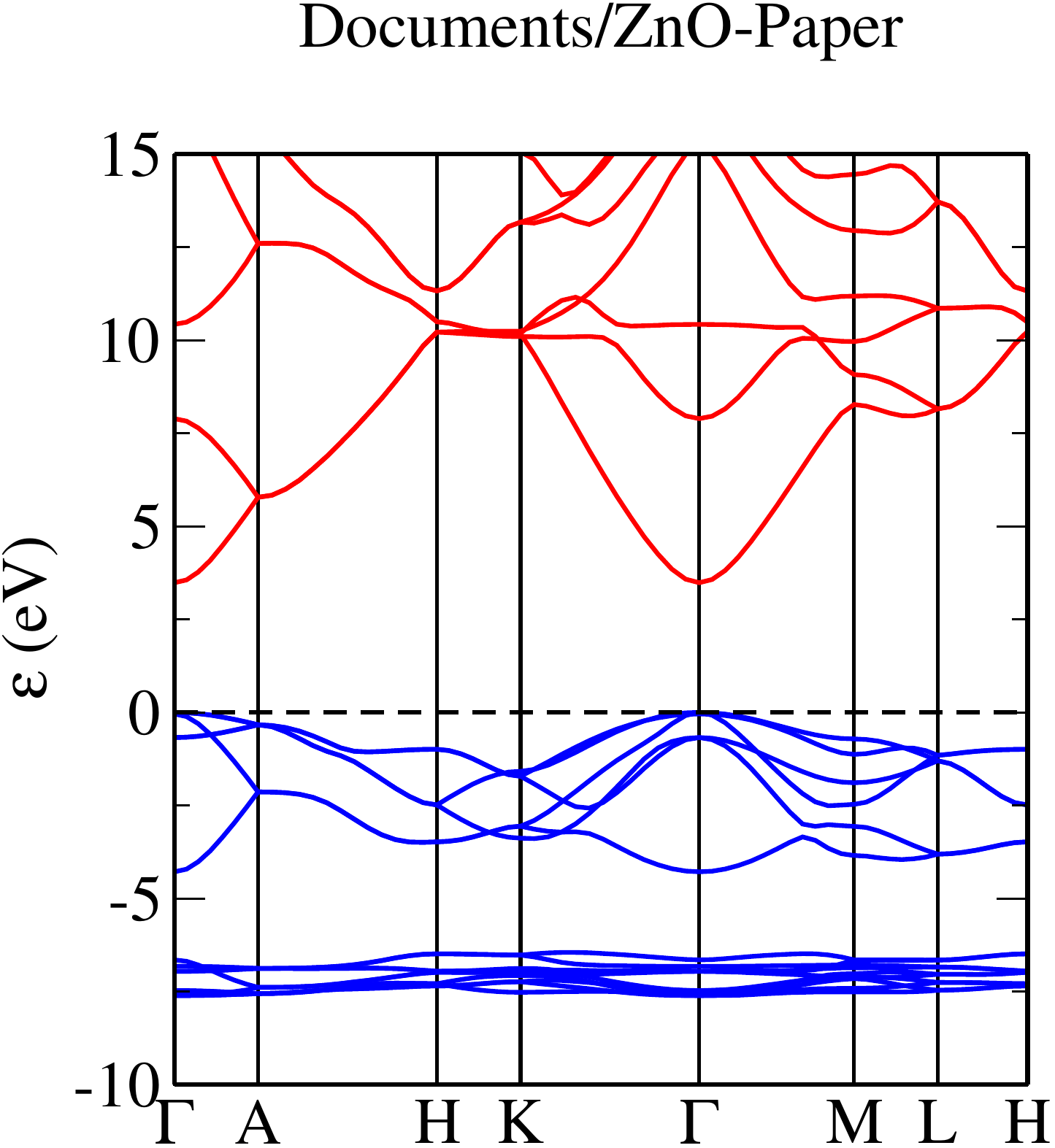}}
\caption{\small Band structure of ZnO (wurtzite structure) using the
  LDA (left) and the OEP (right).}
\label{fig: ZnO}
\end{figure}

The electronic structure of a selection of insulating and
semiconducting materials were calculated using the LDA, some GGAs
(PBE\cite{pbe}, PBESOL\cite{pbe}, PW91\cite{pw91} and WC\cite{wc}), HF
and the OEP. The resultant band gaps are shown
in table~I and for the LDA, OEP and Hartree-Fock
calculations are also plotted in figure~\ref{fig: Linegraph}. The mean
absolute error is 0.55eV for the OEP, 1.39eV for the LDA and 1.19eV
for PBE, although these values are skewed by the underestimated values
for the band gap found for diamond, NaCl and CaO materials. The
corresponding band structures for Ge, CdTe, InN, GaN and ZnO are
plotted in figures \ref{fig: Ge}-\ref{fig: ZnO}.

\begin{table*}
\small
\setlength{\tabcolsep}{10pt}
\begin{tabular}{ l | c c c c c c  c | c }
\hline    
\hline
  & LDA & PBE & PBESOL & PW91 & WC & HF & OEP & Experimental \\ \hline
Ge & 0.09 & 0.13 & 0.04 & 0.14 & 0.04 & 6.01 & 0.86 & 0.79 \cite{bandgapSi} \\
InN & 0.21 & 0.38 & 0.27 & 0.39 & 0.27 & 7.46 & 1.39 & 0.93 \cite{bandgapInN} \\
Si & 0.43 & 0.60 & 0.45 & 0.62 & 0.47 & 6.45 & 1.16 & 1.16 \cite{bandgapSi} \\ 
GaAs & 0.99 & 1.08 & 0.97 & 1.09 & 0.96 & 7.32 & 1.86 & 1.52 \cite{bandgapSi} \\
CdTe & 1.46 & 1.79 & 1.64 & 1.74 & 1.61 & 8.16 & 2.20 & 1.61 \cite{bandgapCdTe} \\
ZnSe & 1.88 & 2.15 & 2.02 & 2.13 & 1.99 & 9.35 & 2.86 & 2.80\cite{bandgapZnSe} \\
GaN & 2.13 & 2.35 & 2.23 & 2.36 & 2.23 & 10.38 & 3.32 &3.39\cite{bandgapGaN} \\
ZnO & 1.93 & 2.28 & 2.09 & 2.29 & 2.10 & 11.51 & 3.48 & 3.43\cite{bandgapZnO} \\
C & 3.98 & 4.21 & 4.03 & 4.24 & 4.08 & 12.76 & 4.87 & 5.47 \cite{bandgapSi} \\
CaO & 3.93 & 4.11 & 4.01 & 4.12 & 4.02 & 14.63 & 6.09 & 7.09 \cite{bandgapCaO} \\
NaCl & 4.84 & 5.34 & 5.13 & 5.35 & 5.19 & 13.74  & 6.27 & 8.97 \cite{bandgapNaCl}\\
\hline
\hline
\end{tabular}
\caption{\small Energy gaps (in eV) for some semiconductors and
  insulators. Calculated values shown for LDA, PBE, PBESOL, PW91, WC,
  Hartree-Fock and the OEP method.}
\end{table*}

\begin{table*}
\small
\setlength{\tabcolsep}{5pt}
\begin{tabular}{ l | c c c c c | c }
\hline    
\hline
& PP PW \cite{Rinke,StadeleGorling1,StadeleGorling2} & FLAPW\cite{Gorling3,KLINorman} & LMTO-ASA\cite{kotani2,kotani4} & KKR-ASA\cite{kotani2,kotani4} & This Work & Experimental \\ \hline
Ge & 0.94 & 0.89 & 1.57, 1.12 & 1.03 & 0.86 & 0.79 \\
InN & 1.40* &&&& 1.39 & 0.93 \\
Si & 1.14, 1.23 & 1.30 & 1.93, 1.25 & 1.12 & 1.16 & 1.16 \\ 
GaAs & 1.78 & 1.74 &&& 1.86 & 1.52 \\
GaN & 2.76*,3.46*,3.49* &&&&3.32& 3.39 \\
ZnO & 2.34* &&&& 3.48 & 3.43 \\
C & 5.06, 4.90 & 5.20 & 5.12, 4.65 & 4.58 & 4.87 & 5.47 \\
CaO & & & 6.15 & 6.29 & 6.09 & 7.09 \\
NaCl & & 6.3$\dagger$ &&& 6.27 & 8.97 \\
\hline
\hline
\end{tabular}
\caption{\small Comparison of calculated OEP band gaps with previous
  work. The
  basis sets are plane wave pseudopotential (PW PP), full potential
  linearized augmented plane wave (FLAPW), linear muffin tin orbitals
  with the atomic sphere approximation (LMTO-ASA) and
  Korringa-Kohn-Rostoker method in the atomic-sphere approximation
  (KKR-ASA). *These results are calculated for the zinc-blende
  structure. $\dagger$This result is obtained with the self
  interaction corrected functional of Perdew and
  Zunger\cite{perdewzunger} rather than the Hartree-Fock exchange
  functional.}
\end{table*}

Figure \ref{fig: Ge} shows the band structure of Ge in the diamond
structure calculated using the LDA and the OEP. For Ge the OEP
improves the gap from the LDA value of 0.09eV to 0.86eV which is very
close to the experimental gap. Similarly for Si the gap is improved
from 0.43eV using the LDA to 1.16eV using the OEP.  For GaAs the OEP
opens the gap from the LDA value of 0.99eV to 1.86eV. The OEP gap is
greater than the experimental gap by 20\%; a similar overestimation of
the gap by the OEP in GaAs was also observed by Gorling \emph{et
  al}.\cite{StadeleGorling1} Figure \ref{fig: CdTe} shows the band
structures for CdTe calculated using the LDA and the OEP. The OEP gap
is 35\% larger than the experimental gap, opening from 1.46eV for the
LDA to 2.20eV for the OEP.

Figure \ref{fig: InN} shows the band structures for the wurtzite
semiconductor InN calculated using the LDA and the OEP. The calculated
band gap for InN has a small direct gap of 0.21eV for the LDA which is
opened to a gap of 1.39eV for the OEP. Similarly the gap in GaAs and
CdTe is overestimated when compared to the experimental gap.  The band
structures calculated for GaN in the wurtzite structure are shown in
figure \ref{fig: GaN}. The OEP band gap of 3.32eV is very close to the
experimental gap of 3.39eV and greatly improved on the LDA band gap
2.13eV.  The wurtzite structure ZnO band structures calculated using
the LDA and the OEP are shown in figure \ref{fig: ZnO}. The calculated
band gap is greatly improved from 1.93eV when using the LDA to 3.48eV
for the OEP which is very close to the experimental gap of 3.43eV.


For ZnSe in the zinc blende structure the OEP calculated gap of 2.83eV
is remarkably close to the experimental gap and greatly improved over
the LDA value of 1.88eV. However for diamond the OEP gap of 4.87eV
underestimates the experimental value by 12\% but still improves on
the LDA value of 3.98eV, the underestimation by the OEP was again
noted by Gorling \emph{et al}.\cite{StadeleGorling1}

For the wide gap rocksalt structure insulators CaO and NaCl the OEP
band gaps are opened up compared to the LDA band gaps, with the LDA
calculated band gap being 3.93eV for CaO and 4.84eV for NaCl and the
OEP calculated band gap being 6.09eV for CaO and 6.32eV for NaCl. For
CaO the gap is underestimated by 14\% for the OEP and for NaCl the
predicted gap is 30\% smaller for the OEP than the experimental
value. The OEP value for CaO is very similar to that found by Kotani
and Akai\cite{kotani2} and for NaCl is in line with that found by Li
\emph{et al}.\cite{KLINorman}

Table II compares the OEP band gaps obtained here to those obtained by
others using a variety of basis sets. The band gaps
calculated here are very similar to previous values and confirm
the progressive underestimation of the band gap for wide gap
materials.


\begin{table*}
\small
\setlength{\tabcolsep}{10pt}
\begin{tabular}{ l | c | c c | c }
\hline    
\hline
 Material & State & LDA & OEP  & Expt. \\ \hline
Ge & Ge $3d$ & 35.01-35.16 & 31.89-32.01 & 29.1-29.6\cite{Gedstate}  \\
InN & In $4d$ & 16.34-17.22 & 18.38-18.94 & 17.4\cite{InNdstate} \\ 
GaAs & Ga $3d$ & 22.95-23.01 & 20.06-20.21 & 18.60-19.04\cite{GaAsdstate} \\
GaAs & As $3d$ & 47.45-47.46 & 44.45-44.49 & 40.37-41.07\cite{GaAsdstate} \\
CdTe & Cd $4d$ & 11.89-12.30 & 10.16-11.03 & 10.10\cite{CdTedstate} \\
ZnSe & Zn $3d$ & 11.65-11.81 & 9.45-9.74 & 9.2\cite{ZnSedstate} \\
ZnSe & Se $3d$ & 61.02-61.03 & 58.50-58.52  & 55.5\cite{ZnSedstate2} \\
GaN & Ga $3d$ & 21.23-21.55 & 18.30-18.85 & 17.94\cite{GaNdstate} \\
ZnO & Zn $3d$ & 8.88-9.69 & 6.65-7.61 & 7.4\cite{ZnOdstate} \\
\hline
\hline
\end{tabular}
\caption{\small Binding energies (eV) relative to the valence band
  maxima for semi-core electrons calculated using the LDA and OEP.}
\end{table*}

The OEP also improves values for semi-core energy levels
relative to the LDA, as shown in table III. This is to be expected given that
the self-interaction error present in the LDA has the largest effect on the tightly-bound d states, 
and that there is no self-interaction error within the OEP.
In Ge the 3d electrons
occupy states in the range -31.9eV to -32.0eV below the valence
band maximum with the experimental binding energy being determined as 29.1eV to 29.6eV. The LDA puts them
in the range -35.0eV to -35.2eV.  The
experimental In 4d electron binding energy in InN is 17.4eV;
the LDA predicts between -18.4eV and -18.9eV
and the OEP between -16.3eV and
-17.2eV. 

For GaAs the experimental Ga 3d states lie between 18.6eV and
19.0eV below the valence band maximum; the OEP predicts a range of
between -20.0eV and -20.2eV and the LDA between
-22.9eV and -23.0eV. Experiment gives As 3d states at 40.4eV to
41.0eV below the valence band maximum; the OEP puts them
at -44.5eV and the LDA predicts
-47.5eV. In CdTe the Cd 4d electrons lie between,
-10.2eV and -11.0eV for the OEP, -11.9eV and -12.3eV for the LDA,
compared with an experimental value of 10.10eV below the valence band
maximum. The experimental results also place the
Te 5s electrons very close to the Cd 4d states with a binding
energy of 9.2eV below the valence band maximum which the OEP also
appears to predict. For ZnSe the Zn 3d electrons lie in the range
-9.4eV to -9.7eV below the valence band maximum for the OEP, -11.6eV to -11.8eV for the LDA and
9.2eV  from experiment. For the Se 3d electrons the LDA gives energies of
-61.02 compared with  -58.5 eV with the OEP and 55.5eV from experiment.

For GaN the LDA gives the Ga d-electrons lying between -21.2eV and
-21.6eV, the OEP between -18.3eV and
-18.9eV with an experimental value of 17.9eV below the valence band
maximum. Using the OEP for ZnO gives Zn 3d
states in the range -6.6eV to -7.6eV relative to the valence
band maximum compared with an experimental value of 7.4eV and a range of -8.9eV to-9.7eV with the LDA. As can be
seen from the band structures for the above materials in the valence,
the s and p states are almost identical when using the LDA and OEP. It
is the d states which display the greatest difference.

The OEP also improves upon the predicted electronic structure given by
the GGA methods of PBE, PBESOL, PW91 and WC which underestimate the
gaps of the materials investigated here and Hartree-Fock which greatly
overestimates the gaps. Further work on magnetic metal-oxides will be
reported in a future publication.


\section{Conclusions}

A method of obtaining the OEP which treats the local exchange
potential exactly without using a sum over all unoccupied states has
been derived using the Hylleraas variational method and ideas borrowed
from density functional perturbation theory. This allows for the
calculation of the OEP using a variational minimization scheme in real
space. The electronic structure of well known materials with a wide
selection of band gaps have been calculated and the band gaps for
semiconductors are found to be in good agreement with experimental
values, although for the larger band gap materials, diamond, CaO and
NaCl, the calculated band gaps are still underestimated by
10-30\%. Hartree-Fock pseudopotentials were found to give more accurate 
results than LDA pseudopotentials. 
%
%
The absence of self-interaction error within the OEP is manifest in a better description of
semi-core d-states compared to the LDA. Their
energies with respect to the top valence band are much closer to 
experimental spectroscopic measurements than within the LDA.

\begin{acknowledgements}
T.W.H. acknowledges the EPSRC for financial support, the UK national
supercomputing facility (HECToR) and finally UKCP for support under
grant EP/F037481/1.
\end{acknowledgements}


\begin{thebibliography}{999}
\bibitem{GGA-exp}
GGAs decay exponentially except for the B88 GGA
  functional which decays as $1/r^2$.
\end{thebibliography}
\end{document}